\documentclass[prb,preprint]{revtex4-1} 

\usepackage{amsmath}  
\usepackage{amsfonts} 
\usepackage{graphicx} 
\usepackage{color}

\begin{document}

\title{The Apparent (Gravitational) Horizon in Cosmology}

\author{Fulvio Melia}
\email{fmelia@email.arizona.edu}
\affiliation{Department of Physics, the Applied Math Program, and Department of Astronomy,
              The University of Arizona, Tucson, AZ 85721}

\date{\today}
\vskip 0.2in
\begin{abstract}
In general relativity, a gravitational horizon (more commonly known as
the ``apparent horizon") is an imaginary surface beyond which all null geodesics
recede from the observer. The Universe has an apparent (gravitational)
horizon, but unlike its counterpart in the Schwarzschild and Kerr metrics, it
is not static. It may eventually turn into an event horizon---an asymptotically
defined membrane that forever separates causally connected events from those
that are not---depending on the equation of state of the cosmic fluid. In
this paper, we examine how and why an apparent (gravitational) horizon is
manifested in the Friedmann-Robertson-Walker metric, and why it is becoming
so pivotal to our correct interpretation of the cosmological data. We discuss
its observational signature and demonstrate how it alone defines the proper
size of our visible Universe. In so doing, we affirm its physical reality and
its impact on cosmological models.
\end{abstract}

\maketitle 

\section{Introduction} 
The term `horizon' in cosmology is variously used to denote (i) how far particles
could have traveled relative to an observer since the big bang (the `particle horizon'), or
(ii) a two-dimensional surface that forever separates causally connected spacetime events from
others that are not (the `event horizon'), or (iii) any of several other definitions
with a customized application.\cite{Rindler:1956} Each has its purpose,
though they are sometimes applied incorrectly when using an approach based on
conformal diagrams, which are often not as easy to interpret as concepts
described in terms of proper distances and times. Part of the difficulty is also
that some definitions are based on the use of comoving lengths, while others
are written in terms of proper distances. There is no conflict between them,
of course, but as the precision of the data increases, it is
becoming quite evident that one definition, above all else, appears to be
the most relevant to a full appreciation of what the observations are telling
us. We shall highlight this particular measure of distance---the `apparent' (or
gravitational) horizon---for the majority of this paper, but we shall also 
compare it to the other horizons in the discussion section towards the end.

Several authors have previously attempted to describe the nature of cosmological 
expansion and its consequences (e.g., its impact on the redshift of light reaching us 
from distant sources) using a pedagogical approach, though with various measures
of success. Given what we know today, primarily through the extensive database at 
our disposal, some of which we shall discuss shortly, it is safe to say that a correct 
understanding of the key observational features in the cosmos is best developed in the
context of general relativity. Of course, this introduces the importance of coordinate
transformations (and their relevance to how one writes the spacetime metric), the
relevance of proper distances and times, and the emergence of the aforementioned 
horizons due to the finite (and constant) speed of light.

The possibility that a gravitational horizon might exist in the cosmos was
considered by Nemiroff and Patla,\cite{Nemiroff:2008} who discussed it using a
``toy" model based on a Universe dominated by a single, isotropic, stable, static,
perfect-fluid energy, though with various levels of pressure. They concluded that
the Friedmann equations implied a maximum scale length over which this energy
could impact an object gravitationally, but suggested that there is little observational
evidence limiting this ``gravitational horizon" of our local Universe. The topic of
this paper touches on this issue as well, but goes well beyond this early, simple
foray into what is---in truth---a much more elaborate physical process, supported
by an enormous amount of empirical evidence. We shall learn, e.g., that the
gravitational horizon in cosmology coincides with what is commonly referred to
as an ``apparent" horizon in other applications of general relativity. An excellent
discussion on the origin and meaning of apparent horizons may be found in the
recent book by Faraoni.\cite{Faraoni:2015} A simple definition that
serves us well at this early stage is that the apparent horizon separates regions
in which null geodesics approach us, from those in which they recede, as 
measured in terms of the proper distance. Indeed, an early recognition of this
dual designation---apparent versus gravitational---was the subject of 
a paper by Gautreau,\cite{Gautreau:1996}
who employed a pseudo-Newtonian description of gravity using a Schwarzschild-like
curvature spatial coordinate, and showed that light signals reach a maximum
distance along their trajectory through the Universe, and then turn around and
return to some remote origin. As we shall see shortly, this turning point is
closely related to an apparent horizon. We shall argue in this paper that the best 
way to understand features such as this from a pedagogical standpoint is actually to 
invoke and utilize the Birkhoff theorem,\cite{Birkhoff:1923} an important generalization 
of Newtonian theory in the general relativistic framework.

In the context of cosmological horizons, the paper by Ellis and Rothman
\cite{Ellis:1993} was quite useful because, in avoiding unnecessarily complicated
presentations, they made it easy to understand how misconceptions often
arise from the misinterpretation of coordinate-dependent effects. These
authors carefully delineated stationary horizons from apparent horizons, extending in a
clear and pedagogical manner the definitions introduced half a century earlier.\cite{Rindler:1956}

In this paper, we focus our attention on the type of horizon that is not
yet commonly known or invoked in the cosmological context---the `gravitational horizon'
which, as noted, is typically referred to as the `apparent horizon,' e.g., in a 
handful of papers dealing specifically with its role in 
cosmology.\cite{Ben-Dov:2007,Faraoni:2011,Bengtsson:2011,Faraoni:2015} As we 
shall see, this horizon has a radius (the `gravitational radius', $R_{\rm h}$) that 
coincides with the much better known radius of the Hubble sphere. In fact,
the very existence of a Hubble radius is due to the presence of the gravitational 
horizon. The former is simply a manifestation of the latter. It is time-dependent---not
static like its counterpart in the Schwarzschild and Kerr metrics---and may,
or may not, eventually turn into an event horizon in the asymptotic future,
depending on the equation of state of the cosmic fluid. (Later in
this paper, we shall directly compare the particle, event, and gravitational
horizons with each other.) Ironically, though
introduced in ref.~\cite{Melia:2007}, an unrecognized form of $R_{\rm h}$
actually appeared a century ago in de Sitter's \cite{deSitter:1917} own
account of his spacetime metric, but the choice of coordinates for which
$R_{\rm h}$ appears in his metric coefficients eventually fell out of favour
following the introduction of comoving coordinates in the 1920's (principally
by Friedmann,\cite{Friedmann:1923}) which have been used ever since in cosmological
applications. We shall see shortly how these two forms of the metric are related
to each other.

As of today, however, there is still some confusion concerning the properties
of $R_{\rm h}$ and how it affects our cosmological observations. The time-dependent
gravitational horizon need {\it not} be a null surface, but is often confused with
one. It has sometimes been suggested \cite{Davis:2004,vanOirschot:2010,Lewis:2013,Kim:2018}
that sources beyond $R_{\rm h}(t_0)$ are observable today, which is absolutely not
the case.\cite{Bikwa:2012,Melia:2012,Melia:2013} Much of this debate appears
to be due to a confusion between proper and coordinate speeds in general relativity.
Simply put, there is no limit on the coordinate speed, which may exceed the speed
of light $c$, but there is absolutely a limit on the proper (or physical) speed, 
whose determination must include the curvature-dependent metric coefficients. When
this distinction is ignored or unrecognized, it can lead to a possibly alarming 
conclusion that recessional velocities in the cosmos can exceed $c$ even within 
the particle horizon of the observer.\cite{Stuckey:1992} As ref.~\cite{Davis:2001}
explains, this ``superluminal" recession is even claimed on occasion to be a contradiction 
of special relativity, which some rationalize by explaining that the limiting speed $c$ is only
valid within a non-expanding space, while the expansion may create superluminal
motion.\cite{Kaya:2011} The aforementioned distinction between coordinate and
proper speeds completely removes any such ambiguous (and sometimes incorrect)
statements.\cite{Melia:2013} Our development of $R_{\rm h}$ in this paper will
be fully consistent with such fundamental aspects of the metric in general
relativity. 

As we shall see shortly, an indication of the role played by $R_{\rm h}$ 
in our interpretation of the data is suggested by the curious coincidence that 
$R_{\rm h}(t)=ct$, a rather significant constraint given that the rate at which 
$R_{\rm h}$ evolves in time is strongly dependent on which components are present
in the cosmic fluid.\cite{Melia:2003,Melia:2007,MeliaAbdelqader:2009,MeliaShevchuk:2012,Melia:2016,Melia:2017}
Those familiar with black-hole horizons might find this result familiar at first, given
that an observer falling freely towards such an object also sees the event horizon
approaching him at speed $c$ (so that $\dot{R}_{\rm h}\equiv dR_{\rm h}/dt=c$). 
But as we shall clarify in this paper, $R_{\rm h}$ is an apparent horizon, not 
necessarily an event horizon, so its evolution in time depends on the
equation of state of the medium. Our principal goal here is therefore to reiterate 
what the {\it physical meaning} of the Universe's apparent (gravitational) horizon is, 
to explore its properties in more detail and at greater depth than has been attempted 
before, and to elucidate its role in establishing the size of the visible Universe based 
on the subdivision of null geodesics that can, or cannot, physically reach us today 
(at time $t_0$).

Those interested in studying the impact of $R_{\rm h}$ at greater depth
may wish to examine how its use eventually resolved the question of whether
or not cosmological redshift is due to an effect distinct from the better 
known kinematic (or Doppler) and gravitational time dilations. This issue 
has been at the core of a debate between those who claim its origin proves
that space is expanding and those who have attempted to demonstrate that it 
may be calculated without invoking such a poorly understood (and probably unphysical) 
mechanism.\cite{Harrison:1995,Chodorowski:2007,Chodorowski:2011,Baryshev:2008,Bunn:2009,Cook:2009,Gron:2007}
A partial demonstration that an interpretation of cosmological redshift
as due to the expansion of space is problematic was presented earlier by Bunn and
Hogg,\cite{Bunn:2009} Cook and Burns,\cite{Cook:2009} and Gron and Elgaroy,\cite{Gron:2007}
but a complete treatment\cite{Melia:2012red} proving that it is simply a product of both 
the Doppler and gravitational redshifts in an expanding cosmos was finalized
only after the introduction of the gravitational radius $R_{\rm h}$.

In \S~2, we shall briefly introduce the Birkhoff theorem and its corollary, and
explain why it is so impactful in helping us understand the relevance of $R_{\rm h}$
to cosmological theory. We shall describe the Friedmann-Robertson-Walker (FRW)
metric in \S~3, and demonstrate how the concept of a gravitational radius
$R_{\rm h}$ emerges from its various forms (based on different choices of the
coordinates). In \S~4, we shall derive the null geodesic equation,
whose properties (and solutions) we study in \S~5. We shall see in this
section that a comprehensive understanding of how the null geodesics behave
leads to a full appreciation of the role of $R_{\rm h}$, which will also help us
understand how $R_{\rm h}$ is related to the particle and event horizons
in \S~6. Finally, we present our conclusions in \S~7.

\section{The Birkhoff Theorem and its Relevance to Cosmology}
The concept of a gravitational radius is not as easily implemented
in cosmology as it is for, say, a compact star. When matter
is distributed within a confined (more or less) spherical volume
surrounded by near vacuum, it is straightforward to understand how
the gravitational influence of the well-defined mass curves the
surrounding spacetime. An apparent (gravitational) horizon appears for
the star when its radius is small enough for the escape speed
to equal or exceed the speed of light, $c$. The Schwarzschild
and Kerr metrics describing the spacetime surrounding such an
environment are time-independent, so their apparent (gravitational)
horizon is actually an event horizon, and these two terms are often
used interchangeably for such objects.

In the cosmological setting, the observations are telling
us that the Universe is spatially flat,\cite{Planck:2016} meaning that
$k=0$ in Equation~(3) below, so it will expand forever and is therefore
infinite. As observers embedded
within it, we do not easily recognize how the gravitational influence
of the cosmic fluid varies with distance. But the reason $R_{\rm h}$
is as important an ingredient of the cosmological metric as it is for
Schwarzschild or Kerr may be understood in the context of the Birkhoff
theorem \cite{Birkhoff:1923} and its corollary (see also
refs.~\cite{Weinberg:1972,Melia:2007}). The theorem itself was actually
`pre-discovered' by Jebsen a few years earlier,\cite{Jebsen:1921}
though this work was not as well known until recently.

The Birkhoff theorem is a relativistic generalization of a Newtonian
theorem pertaining to spherical mass distributions. It states
that for an isotropic distribution of mass-energy, be it static or
time-dependent, the surrounding spacetime is described by the Schwarzschild
metric. The corollary extends this result in cosmologically important ways,
advancing the argument that, for an isotropic Universe, the spacetime
curvature a proper distance $R$ relative to any given origin of the
coordinates depends solely on the mass-energy content within a sphere
of radius $R$; due to spherical symmetry, the rest of the Universe has
no influence on the metric at that radius. Many find this confusing
because the origin may be placed anywhere in the cosmos, so a radius $R$
for one observer, may be a different radius $R^\prime$ for another. The
bottom line is that only {\it relative} spacetime curvature is relevant in this
context. The gravitational influence felt by a test particle at $R$
is relative to an observer at its corresponding origin. For an observer
with a different set of coordinates, the relative gravitational
influence would, of course, be different. To state this another way,
any two points within a medium with non-zero energy density $\rho$
experience a net acceleration (or deceleration) towards (or away) from
each other, based solely on how much mass-energy is present between
them. This is the reason why the Universe cannot be static, for even
though it may be infinite, local motions are dynamically dependent solely
on local densities. Ironically, Einstein himself missed this point---and
therefore advanced the notion of a steady-state universe---because his
thinking on this subject preceded the work of Birkhoff and Jebsen
in the 1920's.

Now imagine the observer extending his perspective to progressively larger
radii. Eventually, his sphere of radius $R$ will be large enough (for the
given density $\rho$) to create a gravitational horizon. In the next
section, we shall prove this rigorously using the Friedmann-Robertson-Walker
(FRW) metric, and demonstrate---not surprisingly---that the radius at which
this happens, called $R_{\rm h}$ throughout this paper, is simply given
by the Schwarzschild form
\begin{equation}
R_{\rm h}={2GM\over c^2}\;,
\end{equation}
where $M$ is the {\it proper} mass contained within a sphere of proper
radius $R_{\rm h}$, i.e.,
\begin{equation}
M\equiv {4\pi\over 3}R_{\rm h}^3\,{\rho\over c^2}\;.
\end{equation}
As it turns out, this is also known in some quarters as the Misner-Sharp
mass,\cite{Misner:1964} defined in the pioneering work of Misner and
Sharp on the subject of spherical collapse in general relativity, and
sometimes also as the Misner-Sharp-Hernandez mass, to include the later
contribution by Hernandez and Misner.\cite{Hernandez:1966,ProperRadius}
But unlike the situation with
the Schwarzschild and Kerr metrics, the cosmological $R_{\rm h}$ may
depend on time, and a sphere with this radius is therefore not necessarily
an event horizon. It may turn into one in our asymptotic future, depending
on the properties of the cosmic fluid. In either case, however, $R_{\rm h}$
defines a gravitational horizon that, at {\it any} cosmic time $t$, separates
null geodesics approaching us from those receding, as we shall
see more formally in Equation~(32) below.

The mass used in Equation~(1) is not arbitrary, in the sense that
only the Misner-Sharp-Hernandez definition is consistent with the $g_{rr}$
metric coefficient. In a broader context, it is highly non-trivial to identify
the physical mass-energy in a non-asymptotically flat geometry in general
relativity.\cite{Faraoni:2015} With spherical symmetry, however, other
possible definitions, such as the Hawking-Hayward quasilocal mass,\cite{Prain:2016} 
reduce exactly to the Misner-Sharp-Hernandez construct.
A second example is the Brown-York energy, defined as a two dimensional surface
integral of the extrinsic curvature on the two-boundary of a spacelike
hypersurface referenced to flat spacetime.\cite{Chakraborty:2015}

Our derivation of the quantity $R_{\rm h}$, though designed for pedagogy, is
nonetheless fully self-consistent with already established knowledge concerning
apparent horizons in general relativity, particularly with their application to
black-hole systems. An apparent horizon is defined in general, non-spherical,
spacetimes by the subdivision of the congruences of outgoing and ingoing null
geodesics from a compact, orientable surface. For a spherically symmetric
spacetime, these are simply the outgoing and ingoing radial null geodesics
from a two-sphere of symmetry.\cite{Ben-Dov:2007,Faraoni:2011,Bengtsson:2011,Faraoni:2015}
Such a horizon is more practical than stationary event horizons in black-hole
systems because the latter require knowledge of the entire future history of
the spacetime in order to be located. Apparent horizons are often used in
dynamical situations, such as one might encounter when gravitational waves
are generated in black-hole merging events.

Unlike the spacetime surrounding compact objects, however, the FRW metric
is always spherically symmetric, and therefore the Misner-Sharp-Hernandez mass
and apparent horizons are related, as we have found here with our simplified
approach based on the Birkhoff theorem and its corollary. Indeed, with spherical
symmetry, the general definition of an apparent horizon reduces exactly to
Equation~(1).\cite{Faraoni:2011,Faraoni:2015} In other words, the use
of Birkhoff's theorem and its corollary allow us to define a `gravitational
horizon' in cosmology which, however, is clearly identified as being the
`apparent horizon' defined more broadly, even for systems that are not
spherically symmetric. It is therefore appropriate for us to refer to
$R_{\rm h}$ as the radius of the apparent horizon in FRW, though given
its evident physical meaning, we shall continue to use the designations
`apparent' and `gravitational' interchangeably throughout this paper,
often combining them (as in `apparent (gravitational) horizon') when
referring to the two-dimensional surface it defines.

The impact of $R_{\rm h}$ on our observations and interpretation of the data
became quite apparent after the optimization of model parameters in the
standard model, $\Lambda$CDM,\cite{Bennett:2003,Spergel:2003,Ade:2014} 
revealed a gravitational radius $R_{\rm h}(t_0)$ equal to $ct_0$ within 
the measurement error.\cite{Melia:2003,Melia:2007}
This observed equality cannot be a mere `coincidence,' as some have 
suggested.\cite{Kim:2018} Consider, for example, that in the context of 
$\Lambda$CDM---a cosmology that ignores the physical
reality of $R_{\rm h}$---the equality $R_{\rm h}=ct$ can be achieved only once
in the entire (presumably infinite) history of the Universe, making it an
astonishingly unlikely event---in fact, if the Universe's timeline is infinite,
the probability of this happening right now, when we happen to be looking, is {\it zero}.

There may be several possible explanations for the existence of such a constraint,
though the simplest appears to be that $R_{\rm h}$ is always equal to $ct$, in which
case this condition would be realized regardless of when the measurements are made.
The unlikelihood of measuring a Hubble radius equal to $ct_0$ today were these
two quantities not permanently linked in general relativity argues for a paramount 
influence of the apparent (gravitational) horizon in cosmology. In subsequent sections 
of this paper, we shall describe how and why $R_{\rm h}$ is manifested through the 
FRW metric. Finally, we shall learn what the photon geodesics in this spacetime are 
informing us about its geometry.

\section{The Friedmann-Robertson-Walker Metric}
Standard cosmology is based on the Friedmann-Robertson-Walker (FRW) metric for a
spatially homogeneous and isotropic three-dimensional space, expanding or contracting
as a function of time:
\begin{equation}
ds^2=c^2\,dt^2-a^2(t)[dr^2(1-kr^2)^{-1}+
r^2(d\theta^2+\sin^2\theta\,d\phi^2)]\;.
\end{equation}
\vskip 0.1in\noindent
In the coordinates used for this metric, $t$ is the cosmic time, measured by a
comoving observer (and is the same everywhere), $a(t)$ is the expansion factor,
and $r$ is the comoving radius. The geometric factor $k$ is $+1$
for a closed universe, $0$ for a flat universe, and $-1$ for an open universe.

As we examine the various properties of the FRW metric throughout this paper,
we shall see that the coordinates $(ct,r,\theta,\phi)$ represent the
perspective of a {\it free-falling} observer, analogous to---and fully consistent
with---the free-falling observer in the Schwarzschild and Kerr spacetimes.
And just as it is sensible and helpful to
cast the latter in a form relevant to the {\it accelerated} observer as well, e.g.,
one at rest with respect to the source of gravity, it will be very informative for
us to also write the FRW metric in terms of coordinates that may be used to
describe a fixed position relative to the observer in the cosmological context.

The proper radius, $R(t)\equiv a(t)r$, is often used
to express---not the co-moving distance $r$ between two points but,
rather---the changing distance that increases as the Universe expands.
This definition of $R$ is actually a direct consequence of Weyl's postulate,\cite{Weyl:1923}
which holds that in order for the Cosmological principle to be maintained
from one time-slice to the next, no two worldlines can ever cross following
the big bang (aside from local peculiar motion that may exist on top of the
averaged Hubble flow, of course). To satisfy this condition, every
distance in an FRW cosmology must be expressible as the product of
a constant comoving length $r$, and a universal function of time,
$a(t)$, that does not depend on position. In some situations, $R$ is
referred to as the areal radius---the radius of two-spheres of
symmetry---defined in a coordinate-independent way by the relation
$R\equiv \sqrt{A/4\pi}$, where $A$ is the area of the two-sphere
in the symmetry. These two definitions of $R$ are, of course, fully
self-consistent with each other.\cite{Nielsen:2006,Abreu:2010} For
example, in this gauge, apparent horizons are located by the
constraint $g^{RR}=0$, which is equivalent to the condition
$\Phi=0$ in Equations~(9) and (10) below.

For convenience, we shall transform the metric in Equation~(3) using
the definition
\begin{equation}
a(t) = e^{f(t)}\;,
\end{equation}
where $f(t)$ is itself a function only of cosmic time $t$.\cite{MeliaAbdelqader:2009}
And given that current observations indicate a flat universe,\cite{Spergel:2003,Ade:2014} 
we shall assume that $k=0$ throughout this paper. Thus, putting
\begin{equation}
r=Re^{-f}\;,
\end{equation}
it is straightforward to show that Equation~(3) becomes
\begin{equation}
ds^2 = \left[1-\left(\frac{R\dot{f}}{c} \right)^2 \right]\,c^2\,dt^2 +
2 \left(\frac{R\dot{f}}{c}\right) c\,dt\,dR-dR^2-R^2\,d\Omega^2\;,
\end{equation}
whereupon, completing the square, one gets
\begin{equation}
ds^2= \Phi\left[c\,dt + \left(\frac{R\dot{f}}{c} \right)\Phi^{-1}
dR  \right]^2 - \Phi^{-1}{dR^2}-R^2\,d\Omega^2\;,
\end{equation}
where
\begin{equation}
d\Omega^2\equiv d\theta^2+\sin^2\theta\,d\phi^2\;.
\end{equation}
For convenience we have also defined the quantity
\begin{equation}
\Phi\equiv 1-\left(\frac{R\dot{f}}{c} \right)^2\;,
\end{equation}
which appears frequently in the metric coefficients. We shall
rearrange the terms in Equation~(7) in order to present the interval in
a more standard form:
\begin{equation}
ds^2= \Phi\left[1 + \left(\frac{R\dot{f}}{c} \right)\Phi^{-1}
{\dot{R}\over c}  \right]^2c^2\,dt^2 - \Phi^{-1}{dR^2}-R^2\,d\Omega^2\;.
\end{equation}
In this expression, $\dot{R}$ is to be understood as representing the
proper velocity calculated along the worldlines of particular observers,
those who have $t$ as their proper time which, as we shall see below,
turn out to be the comoving observers.

This metric has much in common with that used to derive the Oppenheimer-Volkoff
equations describing the interior structure of a star
\cite{Oppenheimer:1939,Misner:1964} except, of course, that whereas the latter is assumed ab
initio to be static, $\dot{R}$ and $\dot{f}$ are functions of time $t$ in a
cosmological context. In other words, the Universe is expanding---in general,
the FRW metric written in terms of $R$ and $t$ is not static. But there are
six special cases where one additional transformation of the time coordinate
(from $t$ to an observer-dependent time $T$) does in fact render all of the
metric coefficients independent of the new time coordinate 
$T$.\cite{Florides:1980,Melia:2013}. These constitute the (perhaps not 
widely known) FRW metrics with constant spacetime curvature. For reference, 
we point out that the standard model is not a member of this special set.

A physical interpretation of the factor $\dot{f}/c$ and, indeed, the function
$\Phi$ itself, may be found through the use of Birkhoff's theorem and its corollary
which, as we have seen, imply that measurements made by an observer a distance
$R$ from his location (at the origin of his coordinates) are unaffected by
the mass-energy content of the Universe exterior to the shell at $R$.
It is not difficult to show from Equations~(9) and (10) that a threshold
distance scale is reached when $R\rightarrow R_{\rm h}$, where $R_{\rm h}$
is in fact equal to $c/\dot{f}$. To see this, we note that the FRW metric
produces the following equations of motion:
\begin{equation}
H^2\equiv\left(\frac{\dot a}{a}\right)^2=\frac{8\pi G}{3c^2}\rho-\frac{kc^2}{a^2}\;,
\end{equation}
\begin{equation}
\frac{\ddot a}{a}=-\frac{4\pi G}{3c^2}(\rho+3p)\;,
\end{equation}
\begin{equation}
\dot\rho=-3H(\rho+p)\;,
\end{equation}
where an overdot denotes a derivative with respect to $t$, and $\rho$ and $p$
represent, respectively, the total energy density and total pressure in the
comoving frame, assuming the perfect fluid form of the stress-energy tensor.

From Equations~(1) and (2), one sees that
\begin{equation}
R_{\rm h}^2={3c^4\over 8\pi G\rho}
\end{equation}
which (with the flat condition $k=0$) gives simply
\begin{equation}
R_{\rm h}={c\over H}={ca\over \dot{a}}\;.
\end{equation}
That is,
\begin{equation}
R_{\rm h}={c/\dot{f}}\;.
\end{equation}
Equation~(15) is fully self-consistent with its well-known counterpart
in the study of apparent horizons in cosmology, as one may trace with greater
detail in refs.~\cite{Faraoni:2011,Faraoni:2015}. Thus, the FRW metric
(Equation~10) may also be written in the form
\begin{equation}
ds^2= \Phi\left[1 + \left(\frac{R}{R_{\rm h}} \right)\Phi^{-1}
{1\over c}\dot{R}  \right]^2c^2\,dt^2 - \Phi^{-1}{dR^2}-R^2\,d\Omega^2\;,
\end{equation}
in which the function
\begin{equation}
\Phi\equiv 1-\left(\frac{R}{R_{\rm h}} \right)^2
\end{equation}
signals the dependence of the coefficients $g_{tt}$ and $g_{RR}$ on the proximity
of the proper distance $R$ to the gravitational radius $R_{\rm h}$.

\section{Geodesics in FRW}
Let us now consider the worldlines of comoving observers. From Weyl's
postulate, we know that
\begin{equation}
\dot{R}=\dot{a}r\;,
\end{equation}
which quickly and elegantly leads to Hubble's law, since
\begin{equation}
\dot{R}={\dot{a}\over a}R\equiv HR\;,
\end{equation}
in terms of the previously defined Hubble constant $H$.\cite{Hconstant}

Therefore, along a (particle) geodesic, the coefficient $g_{tt}$ in Equation~(17) simplifies to
\begin{equation}
g_{tt}=\Phi\left[1+\left({R\over R_{\rm h}}\right)^2\Phi^{-1}\right]^2\;,
\end{equation}
which further reduces to the even simpler form
\begin{equation}
g_{tt}=\Phi^{-1}\;.
\end{equation}
Thus, the FRW metric for a particle worldline, written in terms of
the cosmic time $t$ and the proper radius $R$, is
\begin{equation}
ds^2=\Phi^{-1}c^2dt^2-\Phi^{-1}dR^2-R^2d\Omega^2\;.
\end{equation}
And further using the fact that in the Hubble flow
\begin{equation}
dR=c\left({R\over R_{\rm h}}\right)dt\;,
\end{equation}
Equation~(23) reduces to the final form,
\begin{equation}
ds^2=c^2dt^2-R^2d\Omega^2\;.
\end{equation}

The observer's metric describing a particle moving radially with the Hubble flow
(i.e., with $\dot{R}=HR$ and $d\Omega=0$), is therefore
\begin{equation}
ds=c\,dt\;,
\end{equation}
which displays the behavior consistent with the original definition of our
coordinates. In particular, the cosmic time $t$ is in fact the proper time measured
in the comoving frame anywhere in the Universe, independent of location $R$.
In this frame, we are free-falling observers, and our clocks must therefore
reveal the local passage of time unhindered by any external gravitational
influence.

However, this situation changes dramatically when we examine the
behavior of the metric applied to a fixed
radius $R=R_0$ with respect to the observer. Those familiar with the Schwarzschild
and Kerr metrics in black-hole systems will recognize this situation as being
analogous to that of an observer maintaining a fixed radius relative to the central
source of gravity. The distinction with the previous case is that,
whereas $R$ was there associated with particles (e.g., galaxies) expanding with
the Hubble flow, we now fix the distance $R_0$ and instead imagine particles moving
through this point. Since our measurements are now no longer made from
the free-falling perspective, we expect by analogy with the Schwarzschild
case that gravitational effects must emerge in the metric.

And indeed they do. Returning to Equation~(17), we have in this case
$dR=0$, so that
\begin{equation}
ds^2=\Phi_0 c^2dt^2-R^2_0d\Omega^2\;,
\end{equation}
and if we again insist on purely radial motion (with $d\Omega^2=0$), then
the metric takes the form
\begin{equation}
ds^2=\Phi_0 c^2dt^2\;,
\end{equation}
where now $\Phi_0\equiv 1-(R_0/R_{\rm h})^2$.
In its elegance and simplicity, this expression reproduces the effects
one would have expected by analogy with Schwarzschild and Kerr, in
which the passage of time at a fixed proper radius $R_0$ is now no longer
the proper time in a local free-falling frame. Instead, for any
finite interval $ds$, $dt\rightarrow\infty$ as $R_0\rightarrow 
R_{\rm h}$, which endows $R_{\rm h}$ with the same kind of
gravitational horizon characteristics normally associated with
the Schwarzschild radius in compact objects.

For null geodesics, the situation is quite different, but still fully
consistent with the better known behavior one finds in black-hole
spacetimes. For a ray of light, $\dot{r}$ cannot be zero, as one
may verify directly from the FRW metric in Equation~(3). The
null condition (i.e., $ds=0$), together with a radial path (and,
as always, a flat Universe with $k=0$), leads to the expression
\begin{equation}
c\,dt=\pm a\,dr\;,
\end{equation}
which clearly implies that along an inwardly propagating radial null
geodesic we must have
\begin{equation}
\dot{r}=-{c\over a}\;.
\end{equation}
Thus, in terms of the proper radius $R_\gamma$, the motion of light relative
to an observer at the origin of the coordinates may be described as
\begin{equation}
{dR_\gamma\over dt}=\dot{a}r_\gamma+a \dot{r}_\gamma\;,
\end{equation}
or
\begin{equation}
{dR_\gamma\over dt}=c\left({R_\gamma\over R_{\rm h }}-1\right)\;.
\end{equation}
In this expression, we have assumed that the ray of light is propagating
towards the origin (hence the negative sign). For an outwardly propagating
ray which, as we shall see shortly, is relevant to the definition of a `particle' horizon,
the minus sign would simply be replaced with $+$. It is trivial to
confirm from Equation~(17) that replacing $\dot{R}$ with $dR_\gamma/dt$
from Equation~(32), and putting $dR=(dR_\gamma/dt)\,dt$, gives exactly
$ds=0$, as required for the radial null geodesic.

This result is hardly surprising, but it demonstrates that regardless of
which set of coordinates we use to write the metric for a particle
geodesic, either with or without gravitational effects, the behavior
of light is always fully consistent with the properties expected of
a null geodesic. However, from Equation~(32), we learn several
new important results. First, $dR_\gamma/dt=0$ when
$R_\gamma=R_{\rm h}$. In other words, the spatial velocity of light
measured in terms of the proper distance per unit cosmic time has
two components that exactly cancel each other at the
gravitational radius. One of these is the propagation of light
measured in the comoving frame, where
\begin{equation}
{dR_{\gamma\;{\rm com}}\over dt}=-c\;,
\end{equation}
while the other is due to the Hubble expansion itself, with
\begin{equation}
{dR_{\gamma\;{\rm Hub}}\over dt}=c\left({R_\gamma\over R_{\rm h}}\right)\;.
\end{equation}
Obviously, given the definition of $R_{\rm h}$, we could have simply
written $dR_{\gamma\;{\rm Hub}}/dt=HR_\gamma$.

\begin{figure}
\vskip 1cm
\centerline{
\includegraphics[angle=0,scale=0.5]{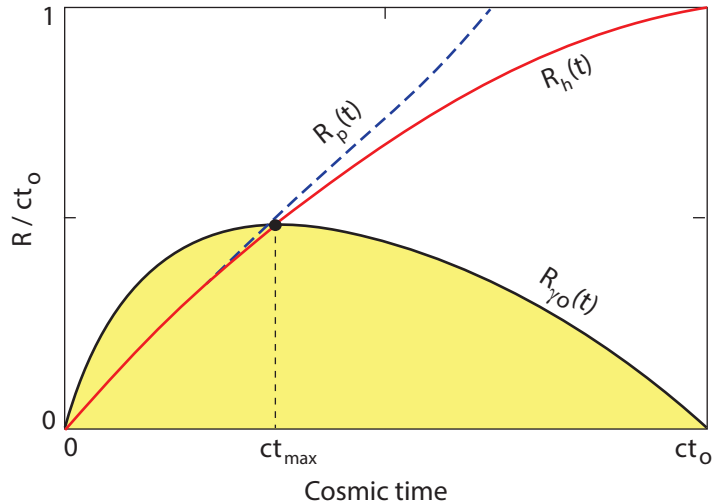}}
\vskip 0.2in
\caption{Schematic diagram showing the various measures of distance discussed
in the context of an apparent (gravitational) horizon. The solid, black curve shows
the null geodesic (i.e., photon trajectory) reaching the observer today, at time $t_0$.
Only photons emitted along this curve can be seen by us at the present time. The solid,
red curve shows the evolution of the gravitational radius $R_{\rm h}(t)$ with time.
(This particular illustration pertains to the standard model, $\Lambda$CDM, though
the qualitative features are common to most cosmological models.) Prior to the time,
$t_{\rm max}$, at which the photon reaches its maximum proper distance relative
to the observer, $R_{\gamma 0}$ is greater than $R_{\rm h}$,
so according to Equation~(32), it must increase with time. This process is reversed
at $t_{\rm max}$, however, after which $R_{\gamma 0}<R_{\rm h}$, so that $\dot{R}_{\gamma 0}<0$. 
Of course, $R_{\gamma 0}\rightarrow 0$ by the time the photon reaches the observer at
$t_0$. In \S~7, we shall also discuss the `particle' horizon, shown here as a dashed
blue curve.}
\end{figure}

Second, we see that when $R_\gamma>R_{\rm h}$, the photon's proper
distance actually increases away from us, even though the photon's velocity
is pointed towards the origin as seen in the co-moving frame (indicated by
the negative sign in Eq.~33). In terms of its proper distance, the photon
approaches us only when it is located within our apparent (gravitational)
horizon at $R_{\rm h}$. Since $R_{\rm h}$ is itself a function of time,
however, ${\dot{R}}_\gamma$ can flip sign depending on whether $R_{\rm h}$
overtakes $R_\gamma$, or vice versa, which may be seen 
schematically in Figure~1. Here, the photon trajectory is represented
by the solid, black curve (labeled $R_{\gamma 0}[t])$, while the gravitational
radius $R_{\rm h}(t)$ is shown in red. For example, a photon emitted beyond
our apparent horizon $R_{\rm h}(t_e)$ at time $t_e<t_0$, corresponding to
the region to the left of $ct_{\rm max}$, may begin its
journey moving away from us (as measured with proper distance), yet stop
when $R_\gamma=R_{\rm h}$, here indicated by the black dot at precisely
$ct_{\rm max}$, and reverse direction at a later time if/when $R_{\rm h}$ will have 
increased faster than $R_\gamma$ and superceded it.
The behavior of the geodesic $R_\gamma$ is therefore heavily dependent
on the cosmology, because the expansion dynamics is solely responsible
for the time evolution of the gravitational radius $R_{\rm h}$. This
behavior of $R_\gamma$, dependent on the value of $R_\gamma/R_{\rm h}$,
affirms the already understood definition of an apparent horizon discussed
in previous work.\cite{Ben-Dov:2007,Faraoni:2011,Bengtsson:2011,Faraoni:2015}

\section{The Apparent (Gravitational) Horizon at $R_{\rm h}$}
This behavior of ${\dot{R}}_\gamma$ confirms the identification of $R_{\rm h}$ as
the radius of a gravitational horizon, albeit an apparent (or evolving) one,
for the observer at the origin of the coordinates. Photons emitted beyond this surface,
even pointing in our direction, actually recede from us, while those emitted
within it follow null geodesics that reach us. In this regard, the apparent
(gravitational) horizon in cosmology behaves like that in the more familiar static
spacetime of Schwarzschild or the stationary spacetime of Kerr but, unlike the
latter, $R_{\rm h}$ is not time-independent in the cosmological setting. Therefore
the horizon may shift, eventually exposing in our future previously unseen
regions at larger proper distances. To be precise, the size of our visible
Universe today, at time $t_0$, hinges on the solution to Equation~(32) for a
given $R_{\rm h}(t)$, starting at the big bang ($t=0$) and ending at the present.
The proper size of our visible Universe is determined by the greatest extent
achieved in proper distance by those null geodesics that actually reach us at
time $t_0$. It is not correct to say that photons we see may cross
$R_{\rm h}$ back and forth without restriction.\cite{vanOirschot:2010,Lewis:2012,Kim:2018}
As we have already noted in the introduction, it is only the null geodesics which
actually reach us that determine the portion of the Universe visible to us
today. A gravitational horizon is observer dependent: light rays that cross
back and forth across the surface at $R_{\rm h}$ and then head to infinity
are completely undetectable by us. And so we arrive at the first constraint
pertaining to the visible Universe:

\hangindent 18pt \textbf{\textit{Constraint I:}} In a cosmology expanding monotonically with
$\dot{H}\le 0$ and ${\dot{R}}_{\rm h}\ge 0$ $\forall\;t\in [0,t_0]$, the proper size of the visible
Universe today is always less than or equal to our gravitational horizon,
i.e., $R_{\gamma,\,{\rm max}}\le R_{\rm h}(t_0)$.
\vskip 0.2in
\hangindent 18pt \textbf{\textit{Proof:}} All null geodesics satisfy the initial boundary
condition $R_\gamma(0)=0$. Null geodesics that reach us must also satisfy
the condition $R_\gamma(t)\rightarrow 0$ as $t\rightarrow t_0$. Therefore,
$\exists$ a time $t_{\rm max}\in [0,t_0]$ (see Figure~1) at which 
$R_\gamma$ has a turning point, i.e.,
\begin{equation}
\left.{dR_\gamma\over dt}\right|_{t_{\rm max}}=0\;.
\end{equation}
\par
\hangindent 18pt According to Equation~(32), this means that
\begin{equation}
R_{\gamma,\,{\rm max}}\equiv R_\gamma(t_{\rm max})=R_{\rm h}(t_{\rm max})\;.
\end{equation}
\par
\hangindent 18pt But ${\dot{R}}_{\rm h}\ge 0$ $\forall\;t\in [0,t_0]$, while
\begin{equation}
{dR_\gamma\over dt}\le 0\quad \forall\; t\in [t_{\max},t_0]\;.
\end{equation}
\par
\hangindent 18pt Therefore $R_{\gamma,\,{\rm max}}=R_{\rm h}(t_{\rm max})\le R_{\rm h}(t_0)$.
\vskip 0.3in

\noindent To illustrate the meaning of this constraint in practice, let us apply
Equation~(32) to two special cases. First, to a cosmology known as the $R_{\rm h}=ct$
universe \cite{Melia:2003,Melia:2007,MeliaAbdelqader:2009,MeliaShevchuk:2012,Melia:2016,Melia:2017}
which, as the name implies, is an FRW-based model with the constraint that $R_{\rm h}$ should always 
be equal to $ct$.  In this cosmology, one has $H(t)=1/t$. Therefore, the null geodesic equation 
becomes
\begin{equation}
{dR_\gamma\over dt}={R_\gamma\over t}-c
\end{equation}
which, together with the boundary conditions $R_\gamma(0)=0$ and
$R_\gamma(t_0)=0$, has the simple solution
\begin{equation}
R_\gamma(t) = ct\,\ln\left({t_0\over t}\right)\;.
\end{equation}
At the turning point, $dR_\gamma/dt=0$, so $t_{\rm max}=t_0/e$, and therefore
\begin{equation}
R^{R_{\rm h}=ct}_{\gamma,\,{\rm max}}={1\over e}\,R_{\rm h}(t_0)\approx 0.37\, R_{\rm h}(t_0)\;.
\end{equation}
In de Sitter space, the Hubble constant $H(t)=H_0$ is fixed, so
the gravitational radius $R^{\rm de\,Sitter}_{\rm h}=c/H_0$ never changes.
In this case, the solution to Equation~(32) is
\begin{equation}
R^{\rm de\,Sitter}_{\gamma,\,{\rm max}}=R_{\rm h}(t_0)\;.
\end{equation}
$\Lambda$CDM falls somewhere in between these two cases. Quite
generally, $R_{\gamma,{\rm max}}$ is typically about half of $R_{\rm h}(t_0)$ for
any given cosmological model (except, of course, for de Sitter).

It has also been suggested that a Universe with phantom energy violates
such observability limits, based on a supposed demonstration that null
geodesics can extend into regions exceeding our gravitational 
horizon.\cite{Lewis:2012,Kim:2018} Aside from the fact that phantom cosmologies
allow for the acausal transfer of energy,\cite{Caldwell:2002,Caldwell:2003}
and are therefore unlikely to be relevant to the real Universe, this
scenario employs null geodesics that never reach us by time $t_0$, and
therefore cannot represent our observable Universe. A second constraint
pertaining to the visible Universe may therefore be posited as follows;

\vskip 0.1in
\hangindent 18pt \textbf{\textit{Constraint II:}} In spite of the fact that
a Universe containing phantom energy (i.e., $p<-\rho$) may have a
gravitational radius $R_{\rm h}(t)$ that changes non-monotonically in
time, none of the null geodesics reaching us today has ever exceeded
our gravitational horizon.
\vskip 0.2in
\hangindent 18pt \textbf{\textit{Proof:}} Let $R^{\,\rm max}_{{\rm h},\,i}$, $i=1...N$,
denote the $N$ rank-ordered maxima of $R_{\rm h}$ on the interval
$t\in [0,t_0]$, such that $R^{\,\rm max}_{{\rm h},\,1}\ge R^{\,\rm max}_{{\rm h},\,2}\ge 
...\ge R^{\,\rm max}_{{\rm h},\,N}$. In addition, let $t_{\rm max}\in [0,t_0]$
be the time at which
\begin{equation}
R_{\rm h}(t_{\rm max})=R^{\,\rm max}_{{\rm h},\,1}\;.
\end{equation}
\par
\hangindent 18pt Now suppose $R_\gamma(t_{\rm max})>R^{\,\rm max}_{{\rm h},\,1}$. In that case,
$dR_\gamma/dt\ge 0$ $\forall\; t\in [t_{\rm max},t_0]$, so $R_\gamma(t_0)\not=0$,
which violates the requirement that photons detected by us today follow null
geodesics reaching us at time $t_0$. In the special case where the Universe
is completely dominated by phantom energy, the horizon is always shrinking
around the observer, so $R_{\rm h}$ has only one maximum (at $t_{\rm max}=0$),
and light rays will reach us only if $R_\gamma<R_{\rm h}$ at the big bang.
\vskip 0.3in

\noindent Together these two constraints make it absolutely clear that no matter
how $R_{\rm h}$ evolves in time, none of the light we detect today has
originated from beyond our gravitational horizon. Indeed, except for de
Sitter, in which the gravitational horizon is pre-existing and static (therefore
leading to Equation~41), all other types of expanding universe have a visibility
limit restricted to about half of our current gravitational radius $R_{\rm h}(t_0)$,
or even somewhat less (e.g., Equation~40). 

\section{Discussion}
As discussed more extensively in ref.~\cite{Melia:2013} for the case of FRW metrics 
with constant spacetime curvature,\cite{Florides:1980} the reason for the 
restriction we have just described is easy to 
understand. In all models other than de Sitter,
there were no pre-existing detectable sources a finite distance from the origin
of the observer's coordinates prior to the big bang (at $t=0$). Therefore,
photons we detect today from the most distant sources could be emitted only
after the latter had sufficient time to reach their farthest detectable proper
distance from us, which is about half of $R_{\rm h}(t_0)$.

The distinction between the `apparent' (gravitational) horizon
$R_{\rm h}$ and the `particle' and `event' horizons in cosmology may now 
be clearly understood. The particle horizon is defined in terms of the maximum
comoving distance a particle can travel from the big bang to cosmic time $t$,
and is given by the solution to Equation~(30) as
\begin{equation}
r_{\rm p}(t) \equiv c\int_0^t {dt^\prime\over a(t^\prime)}\;.
\end{equation}
In terms of the proper distance, the particle horizon is therefore
\begin{equation}
 R_{\rm p}(t) = a(t) c\int_0^t {dt^\prime\over a(t^\prime)}\;.
\end{equation}
If we now differentiate this expression with respect to $t$, we easily
show that
\begin{equation}
\dot{R}_{\rm p}=c\left({R_{\rm p}\over R_{\rm h}}+1\right)\;,
\end{equation}
which needs to be compared with Equation~(32). Given our discussion
in \S~IV, there is no ambiguity about what this equation represents:
it describes the propagation of a photon (i.e., the null geodesic) {\it
away} from the observer at the origin of the coordinates. Its solution
(Eq.~44) therefore gives the maximum proper distance a particle could 
have traveled away from us from the big bang to time $t$. But this is not
the same as the maximum proper distance a photon could have traveled
in reaching us at time $t_0$, given by $R_{\gamma 0}(t_{\rm max})$
in Figure~1, which is always less than $R_{\rm h}(t_0)$, as we have seen.
In contrast, there is no limit to $R_{\rm p}(t)$, since the right-hand side
of Equation~(45) is always greater than $c$, so $R_{\rm p}$ increases
easily past $R_{\rm h}$ (dashed, blue curve in Fig.~1), particularly at late 
times in the context of $\Lambda$CDM, when the cosmological constant 
starts to dominate the energy density $\rho$, and the Universe enters a 
late de Sitter expansion, with both $H$ and $R_{\rm h}$ approaching 
constant values.

So the reason $R_{\rm h}$ is much more relevant to the cosmological 
observations than $R_{\rm p}$, is that we never again see the photons
receding from us, reaching proper distances corresponding to the defined
particle horizon. As we have noted on several occasions, the null geodesics must
actually reach us in order for us to see the photons traveling along them. 

In contrast, the `event' horizon is defined to be the largest comoving distance
from which light emitted now can ever reach us in the future, so the corresponding
proper distance at time $t$ for this quantity is
\begin{equation}
R_{\rm e}(t)\equiv a(t) c\int_t^\infty {dt^\prime\over a(t^\prime)}\;.
\end{equation}
Again differentiating this function with respect to $t$, we find that
\begin{equation}
\dot{R}_{\rm e}=c\left({R_{\rm e}\over R_{\rm h}}-1\right)\;,
\end{equation}
exactly the same as Equation~(32) for $R_{\gamma}$. The physical
meaning of $R_{\rm e}$ is therefore very similar to that of $R_{\gamma}$
except, of course, that by its very definition, the solution in Equation~(46)
represents a horizon for photons that will reach us in our future, not today.
This is why the apparent (gravitational) horizon is not necessarily an event
horizon yet, though it may turn into one, depending on the equation of state in
the cosmic fluid, which influences the solution to Equation~(32).

\section{Conclusions}
We have benefited considerably from writing the FRW metric using two
distinct coordinate systems: (1) the traditional comoving coordinates
that have become very familiar in this context following the pioneering
work of Friedmann in the 1920's, and (2) the coordinates pertaining to
a fixed observer measuring intervals and time at a fixed distance away.
In so doing, we have clearly demonstrated how and why a gravitational
radius is present in cosmology, and how the surface it defines functions
as a horizon separating null geodesics approaching us from those that
are receding. We have proven that all of the light reaching us today originated
from within a volume bounded by this gravitational horizon, clearly
defining the proper size of our visible Universe. 

Going forward, it will be necessary to fully understand why 
$R_{\rm h}(t)$ equals $ct$. A quick inspection of the Friedmann equations
(11-13) suggests that this condition can be maintained if the equation
of state in the cosmic fluid is $\rho+3p=0$, in terms of the total density
$\rho$ and pressure $p$. Why would the Universe have this property?
General relativity makes a distinction between the `passive'  and 
`active' mass: the former is the inertial mass that determines the
acceleration with which an object responds to curvature, while
the latter is the total source of gravity.\cite{Melia:2016,Melia:2017} 
Interestingly, for a perfect fluid in cosmology,\cite{Weinberg:1972} 
the constraint $\rho+3p=0$ means that its active mass is zero, an
elegant, meaningful physical attribute whose consequence is 
zero acceleration, i.e., {\it constant expansion}. Is the Universe really
this simple? If this inference is correct, figuring out why it started 
its evolution with this initial condition will be quite enthralling, to say 
the least.

\begin{acknowledgments}

I am very grateful to Valerio Faraoni for extensive discussions that
have led to numerous improvements to this manuscript. I am also happy
to acknowledge the anonymous referees for providing very thoughtful,
helpful reviews. Some of this work was carried out at Purple Mountain 
Observatory in Nanjing, China, and was partially supported by grant 
2012T1J0011 from The Chinese Academy of Sciences Visiting 
Professorships for Senior International Scientists.

\end{acknowledgments}


\begin{thebibliography}{99}
\bibitem{ProperRadius} {As demonstrated by
these authors, the mass $M(R)$ emerging from the $g_{rr}$ coefficient of
the FRW metric is defined in terms of the comoving density $\rho$ and the
proper radius $R=a(t)r$, where $a(t)$ is the expansion factor and $r$ is
the (unchanging) comoving radius (see Equation~3). Therefore, as one can 
see in Equation~(15), Equations~(1) and (2) lead to an identification
of $R_{\rm h}$ as the Hubble radius only if $R_{\rm h}$ is itself a proper
distance, in contrast to the claim in ref.~\cite{Kim:2018} that
$R_{\rm h}$ is not a proper radius.}
\bibitem{Hconstant}{Of course, $H$ is necessarily ``constant" at every position 
only on a given time slice $t$, but may change with time, depending on the expansion dynamics.}
\end{thebibliography}
\end{document}